\documentclass{PoS}

\title{Recent results in chiral effective field theory for the NN system}

\ShortTitle{Recent results in $\chi$EFT for NN}

\author{\speaker{Daniel R.~Phillips}\\
        Department of Physics and Astronomy and Institute of Nuclear and Particle Physics, Ohio University, Athens, OH 45701, USA\\
        E-mail: \email{phillips@phy.ohiou.edu}}

\abstract{I describe recent progress towards a theory of the NN force which captures the consequences of QCD's chiral symmetry and the pattern of its breaking, and is formulated as an expansion in a ratio of low and high mass scales, $M_{\rm lo}/M_{\rm hi}$. This ``chiral effective field theory" of the NN system is a firm foundation for explorations of nuclear structure and reactions that are grounded in QCD's low-energy symmetries. While calculations that use a $\chi$PT expansion for the NN potential have proven very successful, they can only be used with a narrow range of momentum-space cutoffs, which leaves the expansion parameter  for observable quantities somewhat murky. Here we seek a truly systematic effective field theory for the NN amplitude, that is manifestly renormalization-group invariant at each order in a demonstrably perturbative expansion.}

\FullConference{The 7th International Workshop on Chiral Dynamics,\\
		August 6 -10, 2012\\
		Jefferson Lab, Newport News, Virginia, USA}

\begin{document}

\section{$\chi$EFT for the NN system since Weinberg: what is an appropriate level of naivet\'e?}

This series of meetings focuses on manifestations of chiral dynamics in systems of hadrons. Over the past twenty years, effective field theories (EFTs) have played a key role in improving our knowledge in this area. In particular, chiral perturbation theory ($\chi$PT) has been applied, with great success, to meson dynamics, and to the interactions of pions and photons with a single nucleon. Nuclei, however, cannot be directly addressed within $\chi$PT. $\chi$PT yields a purely perturbative expansion in powers of $(p,m_\pi)/\Lambda_{\chi {\rm SB}}$ (with $\Lambda_{\chi {\rm SB}}$ nominally $m_\rho \sim 4 \pi f_\pi$) for the scattering amplitude. Nuclei are bound states, and so cannot be generated from such an expansion. In this talk I will describe twenty years of effort to develop an EFT, which I refer to as $\chi$EFT, that extends the benefits of $\chi$PT to systems with $A \geq 2$. That EFT should have the following features:
\begin{itemize}
\item Encode the consequences of QCD's spontaneously and explicitly broken chiral symmetry.
\item Treat at least part of the problem non-perturbatively, so that bound states emerge naturally.
\item Nevertheless, yield predictions for observables which can be understood as an expansion in $M_{\rm lo}/M_{\rm hi}$ (with $M_{\rm lo}$ and $M_{\rm hi}$ mass scales ``to be named later").
\item Be renormalizable order-by-order in this expansion parameter.
\end{itemize}
Such a treatment of few-nucleon systems would enable {\it a priori} error estimates. It would also permit identification of features of those systems which are a consequence of the pattern of chiral-symmetry breaking in QCD, thereby linking nuclear physics to QCD in a model-independent way. 

In the early 1990s Weinberg argued that the infrared enhancement associated with multi-nucleon intermediate states meant that the $\chi$PT expansion could not be applied directly to the scattering amplitude in systems with $A > 1$~\cite{We90}. He
pointed out that the $\chi$PT Lagrangian and counting rules {\it could} be used to compute the NN potential, $V$ (and, by extension, any two-nucleon-irreducible NN operator), up to some fixed order, $n$, in $\chi$PT. Such an expansion can then be examined for convergence with $n$.
At the energies where $\chi$PT can usefully be applied to nuclear physics the nucleons are non-relativistic, and so Weinberg proposed to incorporate the effects of NN intermediate states in $\chi$EFT by inserting the $\chi$PT $V$ into the Schr\"odinger equation:
\begin{equation}
(E - H_0)|\psi \rangle=V|\psi \rangle,
\label{eq:SE}
\end{equation}
with $|\psi \rangle$ the wave function of the NN system. 
At leading order (LO) the potential may be written:
\begin{equation}
\langle {\bf p}'|V^{(0)}|{\bf p} \rangle=C^{3S1} P_{3S1} + C^{1S0} P_{1S0} + V_{1 \pi}({\bf p}'-{\bf p}); \quad V_{1 \pi}({\bf q})=-\tau_1^a \tau_2^a \frac{g_A^2}{4 f_\pi^2}\frac{\sigma_1 \cdot {\bf q} \sigma_2 \cdot {\bf q}}{{\bf q}^2 + m_\pi^2},
\label{eq:V0}
\end{equation}
where $V_{1 \pi}$ is the one-pion-exchange potential, $P_a$ is a projector onto the NN partial wave $a$, and $C^a$ is a low-energy constant (LEC) appearing in the NN piece of the chiral Lagrangian. These are the only two contact operators in the NN part of the chiral Lagrangian which have dimension zero. Thus, according to naive dimensional analysis (NDA), they should be the only ones present in this chiral-dimension-zero operator. To solve Eq.~(\ref{eq:SE}) with this $V$  one must introduce a cutoff, $\Lambda$, on the intermediate states, because  the $\chi$PT potentials grow with momenta. The contact interactions should then absorb the dependence of the effective theory's predictions on $\Lambda$ in all low-energy observables. Otherwise we conclude that the theory is unable to give reliable predictions. 

The $\chi$PT potential $V$ was computed to $O(P^3)$ in Refs.~\cite{Or96,Ep99,EM02}, and to $O(P^4)$ in Refs.~\cite{EM03,Ep05}. The various NN LECs that appear in $V$ were fitted to data for $\Lambda$ in the range 500--800 MeV. The resulting predictions---especially the ones obtained with the $O(P^4)$ potential---contain very little residual cutoff dependence in this range of $\Lambda$'s, and describe NN data with considerable accuracy. 

Consistent three-nucleon forces have been derived and implemented in such an approach~\cite{vK94,Ep02}. The NN and 3N forces have been employed to compute nuclei up to sizable values of $A$, with significant gains in our understanding of nuclear structure~\cite{MVN12,Ro12}. They have also been used to compute 3N scattering~\cite{KN11}, and are now being employed to describe more complicated scattering processes~\cite{NQ11}. The theory of few-nucleon systems proposed by Weinberg has also been used for 
electron-deuteron scattering~\cite{Ko12}, magnetic moments and capture in light nuclei~\cite{Pa12}, Helium-4 photoabsorption~\cite{QN07}, weak reactions~\cite{Ma12}, and Compton scattering in the NN and NNN systems~\cite{Gr12}. The approach implemented in these works, and many others, has been very successful, leading to marked improvements in the phenomenology of few-nucleon systems, and insights into 3N forces which are relevant to larger nuclei, nuclear matter and neutron stars.

But, a nagging question remains: is it an EFT? While there is an expansion for $V$, that $V$ is then iterated to all orders via Eq.~(\ref{eq:SE}), so there is no---or at least, no obvious---perturbative expansion for observables. Moreover, if ``renormalization" is taken to have its usual meaning in an EFT,  this approach is not properly renormalized. Several papers have shown that Eqs.~(\ref{eq:SE}) and (\ref{eq:V0}) do not yield stable predictions once cutoffs larger than $\sim 800$ MeV are considered~\cite{Be02,ES03,NTvK,Bi06,PVRA06,Ya08}. The impact of short-distance physics on observables is thus not under control. In an EFT taking the cutoff $\Lambda \rightarrow \infty$ does not necessarily result in a decrease in the errors due to short-distance physics in the calculation~\cite{Le97,EM06}. But, at the very least, we should  be able to vary $\Lambda$ by a factor $\sim 2$ around $\Lambda_{\chi {\rm SB}}$, without the predictions of our ``$\chi$EFT" going haywire. This criterion is not satisfied if $V^{(0)}$---or higher-order $V$'s~\cite{Ya09A,Ya09B}---are inserted in Eq.~(\ref{eq:SE}). The two connected deficiencies of $\chi$EFT as presently practiced: lack of an expansion parameter and lack of renormalization, make it quite difficult to to formulate {\it a priori} error estimates in a calculation.

In order to obtain a theory which can truly be called ``$\chi$EFT" we need to re-examine the reason for iterating $V^{(0)}$, so as to obtain a well-defined, renormalized leading order. We can then perturb around this result, computing higher orders in the EFT which are demonstrably small in an expansion parameter $M_{\rm lo}/M_{\rm hi}$. 
I emphasize that I am {\it not} suggesting that all of nuclear physics should be calculated in this perturbative expansion. This is not necessary, or even wise. The goal is to understand what terms must be included in order to achieve some specified accuracy $(M_{\rm lo}/M_{\rm hi})^n$. We can then include that physics in a potential which can be used with a low cutoff to calculate nuclear structure and reactions. In this way we may {\it a postieri} justify (or even, perhaps, improve upon!) the work that has been done using $\chi$PT expansions for $V$ and other operators. 

\section{The ``new leading order" and its discontents}

\label{sec-newLO}

The first step towards this goal is to understand the size of iterates of one-pion exchange.  Iterates of $V_{1\pi}$ become comparable with the tree-level graph for momenta of order
$
\Lambda_{NN}=\frac{16 \pi f_\pi^2}{g_A^2 M} \approx 300~{\rm MeV}$~\cite{Ka98,FMS,Be02,Bi06}.
Crudely speaking, $\Lambda_{NN}$ is the momentum scale at which the tensor part of one-pion exchange becomes non-perturbative. That part of $V_{1 \pi}({\bf r})$ corresponds to a $1/r^3$ potential that couples NN partial waves obeying $\Delta L=2$. Importantly, the tensor part of one-pion exchange does not appear in waves with $S=0$. It is also screened by the centrifugal barrier in partial waves of sufficiently large $L$~\cite{Bi06}. 

Consequently one-pion exchange {\it can} be treated perturbatively in higher partial waves, as has been known for decades. $V_{1 \pi}$ is weak enough that employing the standard $\chi$PT counting rules to compute the NN scattering amplitude yields a good description of NN phase shifts in F-, G-, $\ldots$ waves, as demonstrated by Kaiser et al. more than 15 years ago~\cite{Ka97}. 

In contrast, in partial waves of low $L$ with $S=1$, we need to iterate one-pion exchange if we wish to describe processes at $p \sim \Lambda_{NN}$. In this regime leading order requires that we treat $V_{1 \pi}$ non-perturbatively, i.e. solve the Schr\"odinger equation with this potential. In the ${}^3$S$_1$-${}^3$D$_1$ channel, in the triplet P-waves, and possibly also in D-waves, we must generate the full LO Green's function corresponding to Eqs.~(\ref{eq:SE}) and (\ref{eq:V0}) as:
\begin{equation}
G=G_0 + G_0 V^{(0)} G,
\label{eq:GF}
\end{equation}
with $G_0$ the free Green's function for the propagation of two nucleons. 
This begins to define LO in an NN $\chi$EFT with $M_{\rm lo}=\{m_\pi,p,\Lambda_{NN}\}$ and $M_{\rm hi}=\Lambda_{\chi {\rm SB}}$. 
 Higher-order pieces of the $\chi$PT potential---most notably the two-pion-exchange graphs in $V^{(2)}$ and $V^{(3)}$---can be treated as perturbations, since they are suppressed by 
 $(\Lambda_{NN}/\Lambda_{\chi {\rm SB}})^n$.
In order to make sure the LO calculation is renormalized $V_{1 \pi}$ needs to be supplemented by contact interactions. 
The issue that must be resolved is: which ones? If the two written in Eq.~(\ref{eq:V0}) are not sufficient, what guide shall we use to decide which contact operators that are higher order from the point of view of NDA should be included?

\begin{figure}[h]
\vspace*{-1.3in}
\centerline{
\includegraphics[width=4in]{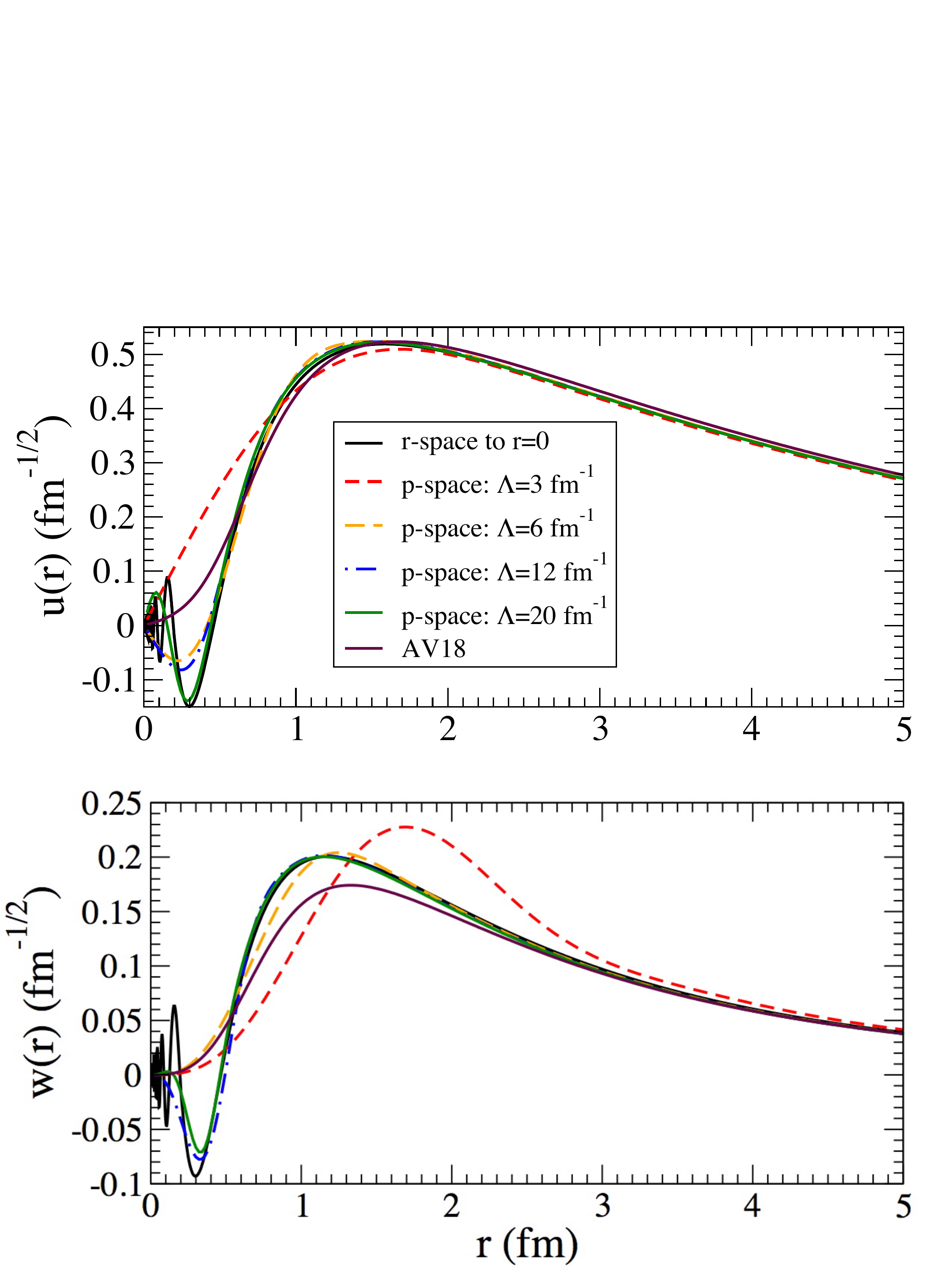}}
\vspace*{-0.15in}
\caption{Predictions for the deuteron radial wave functions $u(r)$ ($^3$S$_1$) and $w(r)$ ($^3$D$_1$) for different values of the momentum-space cutoff $\Lambda$. Also shown, for comparison, are the radial wave functions of the AV18 potential~\cite{PV08}.}
\label{fig-deutwfs}
\end{figure}

In S waves the two contact interactions of Eq.~(\ref{eq:V0}) {\it are} sufficient, and the solution to Eq.~(\ref{eq:SE}) is stable for a wide range of cutoffs~\cite{Be02,PVRA05,Ya08}. From the Green's function we extract wave functions for the deuteron (see Fig.~\ref{fig-deutwfs}) that can be used to compute, e.g. electron-deuteron scattering~\cite{PV08}.

Features of these wave functions reflect the presence of the tensor $1/r^3$ potential. At short distances this singular, attractive potential generates two regular solutions:
\begin{equation}
u_1^A(r)=(\Lambda_{NN} r)^{3/4} \cos\left(4 [\Lambda_{NN} r]^{-1/2}\right); u_2^A(r)=(\Lambda_{NN} r)^{3/4} \sin\left(4 [\Lambda_{NN} r]^{-1/2}\right).
\label{eq:uA}
\end{equation}
The Schr\"odinger equation with a potential $\sim 1/r^n$ with $n < 2$ has one regular and one irregular solution. In contrast, here we have two equally regular solutions. A boundary condition is needed to fix their relative phase~\cite{Case50,Sprung94,Be01,Bi06,PVRA06}. A repulsive $1/r^3$ potential generates solutions:
\begin{equation}
u_1^R(r)=(\Lambda_{NN} r)^{3/4} \exp\left(-4 [\Lambda_{NN} r]^{-1/2}\right); u_2^R(r)=(\Lambda_{NN} r)^{3/4} \exp\left(4 [\Lambda_{NN} r]^{-1/2}\right),
\label{eq:uR}
\end{equation}
and, here too, in principle, a boundary condition is required to set the relative strength at short distances. However, the choice of that boundary condition is numerically irrelevant, because of the rapid exponential fall off of $u_1^R$.

\begin{figure}[htbp]
\vspace*{-0.1in}
\centerline{
\includegraphics[width=3in]{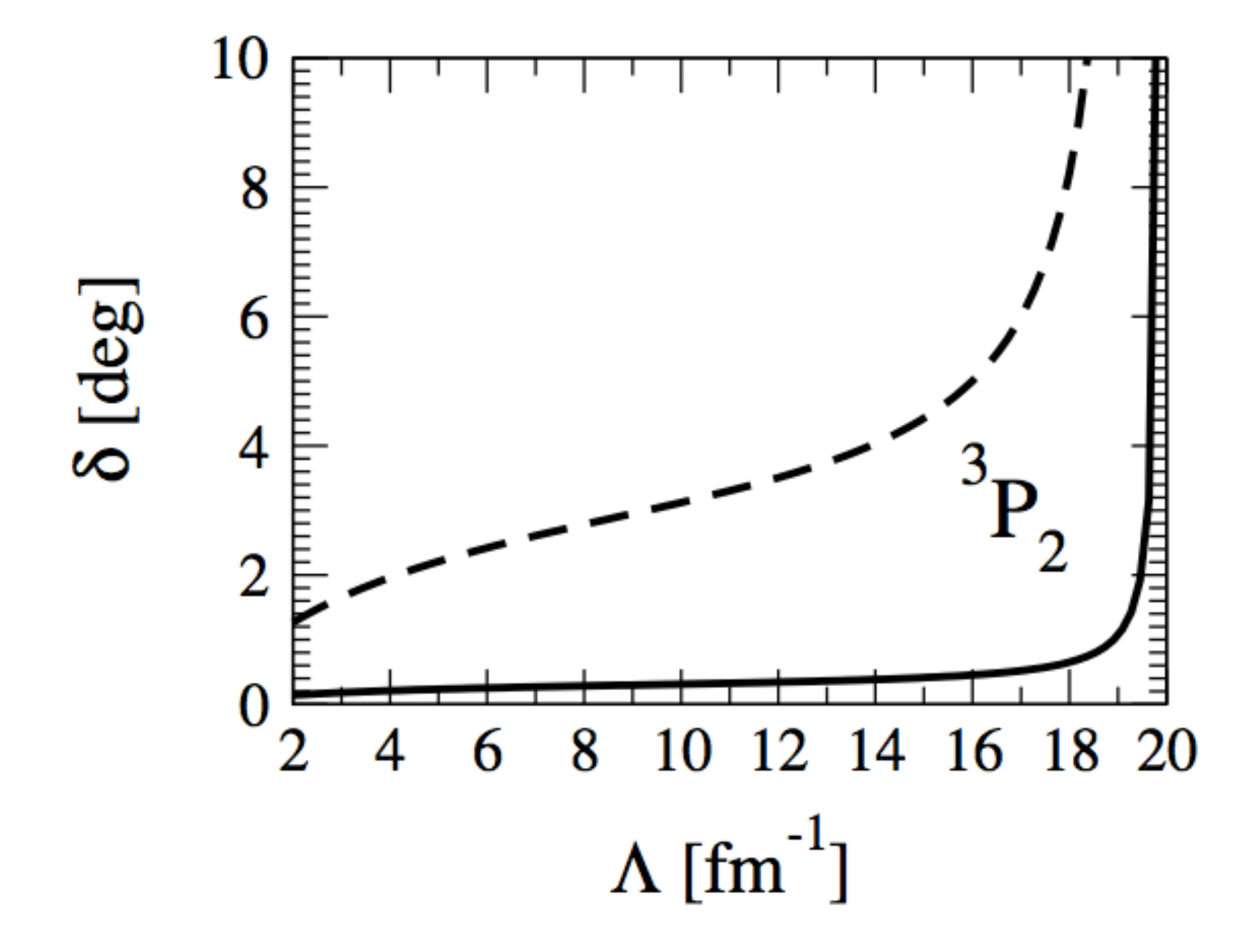}}
\vspace*{-0.2in}
\caption{Phase shifts at $T_{\rm lab}=10$ MeV (solid) and 50 MeV (dashed), as a function of the momentum-space cutoff $\Lambda$, for the NN ${}^3$P$_2$ channel with $V=V_{1 \pi}$. From Ref.~\cite{NTvK}. Used with permission.}
\label{fig-PwavephasesLO}
\end{figure}

This behavior of wave functions at distances $r \ll 1/\Lambda_{NN}$ is, it turns out, what drives the cutoff dependence and independence of calculations with $V^{(0)}$ in different partial waves. In the ${}^3$S$_1$-${}^3$D$_1$ case, the contact term $\sim P_{3S1}$ in Eq.~(\ref{eq:V0}) sets the phase between $u_1$ and $u_2$. In consequence the predictions for deuteron wave functions $u(r)$ and $w(r)$ are stable as $\Lambda \rightarrow \infty$, as seen in Fig.~\ref{fig-deutwfs}. But, if $V^{(0)}$ is used to compute, e.g. phase shifts, in a channel where such a contact interaction is not present, there is no mechanism that enforce a particular short-distance phase between $u_1^A$ and $u_2^A$. That phase is instead set by the regulator at $p \sim \Lambda$, and this regulator dependence propagates to long distances and yields phase shifts that are strongly cutoff dependent, see, e.g. Fig.~\ref{fig-PwavephasesLO}.

This answers the question of how to supplement $V_{1 \pi}$ by contact interactions and obtain a renormalized LO $\chi$EFT calculation. 
As advocated in Ref.~\cite{NTvK}, contact terms must be included that set the phase at short distances when that phase is relevant. This means adding to $V^{(0)}$, at least, contact terms $\sim P_{3P0}$ and $\sim P_{3P2}$. This yields a ``new leading order" $V$ of $\chi$EFT. (I say more about whether contact operators must be included in D-waves below.) There appears to be no need to include a LO contact term $\sim P_{3P1}$, since the tensor piece of one-pion exchange is repulsive there, and so $V_{1\pi}$ can be iterated to generate cutoff-independent predictions. 
The qualitative arguments I present here regarding the correct LO $V$ for $\chi$EFT can be made rigorous by formulating the problem as one of requiring renormalization-group (RG) invariance for, e.g. $G$ of Eq.~(\ref{eq:GF}), see Ref.~\cite{Bi06}. 

Regardless of whether the argument is made in momentum space or co-ordinate space, and whether it is made using the RG or by examining the solutions of the Schr\"odinger equation, the conclusion is that the NN-system wave function is highly distorted at short distances, because of the strong $1/r^3$ potential that is operative there. The failure of naive dimensional analysis to predict which contact operators are necessary for renormalization at leading order is just a consequence of the fact that these solutions are {\it not} plane waves, and are not even perturbatively close to being plane waves. The scaling of short-distance operators is thus very different from what it is in $\chi$PT. 

Various attempts to circumvent these complications due to strong interactions at short distances have been suggested. Beane, Kaplan, and Vuorinen proposed to make $V_{1 \pi}$ weaker, by applying a Pauli-Villars regulator. They argue it can even be made perturbative, with its short-distance strength shuffled into contact operators. This scheme has been worked out up to NNLO for the ${}^3$S$_1$-${}^3$D$_1$ partial wave, with encouraging results~\cite{BKV}. Its performance in the controversial P-waves has not been publicized. Recent work described by Gegelia at this meeting has proposed the use of a relativistic NN propagator for $G_0$~\cite{GE12}. This softens the integrals for iterates of $V_{1 \pi}$. Indeed, at short distances the problem can be shown to be the solution of a $1/r^2$ potential in two dimensions. Some contact terms beyond those mandated by NDA are still needed, but not as many as were proposed in Ref.~\cite{NTvK}. Lastly, Epelbaum and Mei\ss ner argued that discussing the behavior of $V_{1 \pi}$ for $r \ll 1/\Lambda_{NN}$ is a fruitless exercise, since the small separation between $\Lambda_{NN}$ and $\Lambda_{\chi {\rm SB}}$ means that we very quickly descend into an examination of features of the $\chi$PT potential which are artifacts of $\chi$PT's status as an effective theory~\cite{EM06}. For this reason, many practitioners continue to advocate, and employ, a cutoff which never rises above values $\sim m_\rho$. 

At least two of these suggestions could lead to a systematic $\chi$EFT for few-nucleon systems. This contribution is not long enough that I can do all the proposals justice, so in the space that remains I will discuss higher orders in the EFT that is built upon the ``new leading order" proposed in Ref.~\cite{NTvK}. This is also the proposed expansion for which the most detailed analyses of higher-order effects have been carried out.

\section{Higher orders in $\chi$EFT}

\label{sec-subleading}

For an EFT to be useful it must converge as higher orders in the expansion are computed. In this section we discuss the behavior of a perturbative expansion built on the LO calculation defined in the previous section.

The EFT expansion for the long-distance part of the $\chi$PT potential $V$ is not controversial. Once the particle content of the theory is fixed---and here we deal solely with the ``Delta-less" theory that contains only pion and nucleon degrees of freedom---the relative size of the various two-pion, three-pion, etc. mechanisms is dictated by the power counting of $\chi$PT, as envisioned by Weinberg~\cite{We90}. The contentious issue is the power counting for the short-distance operators.  We will now show how power counting for these operators can be established via a renormalization-group analysis of matrix elements of the long-distance potential between LO wave functions.

\begin{figure}[h]
\centerline{
\includegraphics[width=3in]{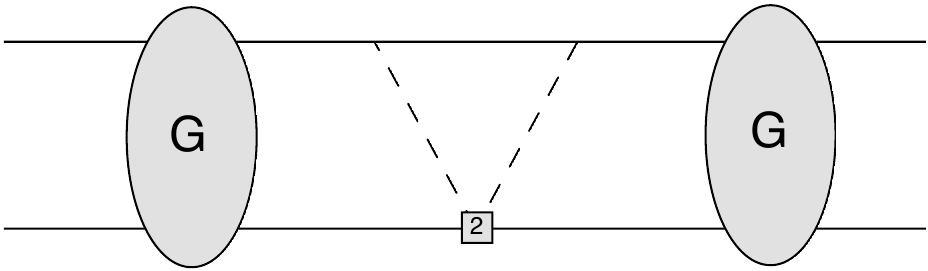}}
\vspace*{-0.15in}
\caption{Diagram representing the perturbative correction to the leading-order Green's function due to an insertion of the sub-leading two-pion-exchange potential derived from $\chi$PT.} 
\label{fig-correction}
\end{figure}

We take as an example matrix elements of the ``sub-leading two-pion-exchange potential", $V_{2 \pi}^{(3)}$. This generates the correction to the leading-order Green's function $G$ depicted in Fig.~\ref{fig-correction}.
The pertinent matrix elements of this potential are $\langle \psi^{(0)}|V_{2 \pi}^{(3)}|\psi^{(0)} \rangle$, where $|\psi^{(0)} \rangle$ is the wave function extracted from the (renormalized) LO Green's function (\ref{eq:GF}). We will add to this a matrix element of a sub-leading contact interaction (which must also be inserted between LO $G$'s), in order to obtain an overall finite result. The contact interaction must be chosen such that the total result is independent of the short-distance cutoff placed on the integrals in the matrix-element calculation, $r_c$. To compute the dependence of the matrix element of $V^{(3)}$ on $r_c$ we need only the short-distance behavior of the LO wave functions (\ref{eq:uA}) and (\ref{eq:uR}) and that  $M V_{2 \pi}^{(3)} \sim \frac{\Lambda_{NN}}{\Lambda_{\chi {\rm SB}}^4 r^6}$. It then follows that 
\begin{equation}
\langle \psi^{(0)}|M V_{2 \pi}^{(3)}|\psi^{(0)} \rangle \sim \frac{\Lambda_{NN}^{5/2}}{\Lambda_{\chi {\rm SB}}^4} \frac{1}{r_c^{7/2}} + \frac{\Lambda_{NN}^{5/2}}{\Lambda_{\chi {\rm SB}}^4} \frac{k^2}{r_c^{3/2}} + \mbox{positive powers of $r_c$}.
\label{eq:poormansRG}
\end{equation}
Note that the expansion of the short-distance wave functions as a function of the NN system energy $E=k^2/M$~\cite{PVRA06} shows that this is an expansion in $k^2$.

Eq.~(\ref{eq:poormansRG}) implies that two contact interactions---one a constant and one proportional to $E$---are needed for renormalization at this order. (The fact that the contact interactions must themselves be sandwiched between $|\psi^{(0)} \rangle$'s  does not alter this conclusion.) In S-waves this is the same result as found by naive dimensional analysis, although the scaling of the contact operators turns out to be different. The real difference occurs in P-waves, where the $r^2$ behavior of plane waves at short distances which would be assumed in NDA is replaced by $r^{3/4}$ dependence because of the singular LO potential. Therefore two contact interactions are also needed to renormalize $V_{2 \pi}^{(3)}$ in P-waves, and, indeed, in any partial wave where $V_{1\pi}$ is attractive and sufficiently strong to warrant iteration.

One channel where we know non-perturbative effects are present, but $V_{1 \pi}$ is not strong, is the ${}^1$S$_0$. The tensor part of $V_{1 \pi}$ plays no role there. In consequence, all indications are that one-pion exchange can be treated as a perturbation in this channel~\cite{Ka98,FMS,Be02,Sh08}. In $\chi$EFT the unnaturally large scattering length, $a$, in this channel results from fine-tuning of the contact interaction to be $O(P^{-1})$. This produces a LO wave function that behaves as $a/r$ at short distances. Matrix elements of higher-order pieces of the chiral potential are then highly divergent~\cite{Bi10,PV09,PV11,LY12}. The sub-leading contact terms $C_2^{1S0} k^2$, $C_4^{1S0} k^4$, etc. must be enhanced compared to their NDA order to absorb these divergences. The resulting power counting is very similar to that of the ``pionless EFT"~\cite{Ka98,vK99,Ge98,Bi99}, but long-distance corrections due to pion exchanges are included in $\chi$EFT. Since the ${}^3$S$_1$ also has a large (c.f. $1/m_\pi$) scattering length there is a similar, but not identical, enhancement of sub-leading contact interactions there~\cite{Bi09}.

\section{Results}

The formal outcome of this kind of RG analysis was summarized by Birse at the last Chiral Dynamics meeting~\cite{Bi06,Bi09}. I reproduce his table  below. The progress since 2009 has largely been in implementing this understanding of how the different pieces of the $\chi$EFT $V$ should be ordered in calculations, and comparing the results with data. We now turn our attention to highlights from those investigations. More details can be found in the original papers~\cite{PV09,PV11,LY11A,LY11B,LY12} and in the contributions of Long and Pavon Valderrama to this meeting~\cite{Long,Pavon}.

\begin{table}[h]
\begin{center}
\begin{tabular}{|c|c|}
\hline
$P^{-1}$ & $V_{1\pi}$, $C^{3S1}$, $C^{1S0}$\\
$P^{-1/2}$ & $C^{3P0}$, $C^{3P2}$\\
$P^0$ & $C_2^{1S0}$\\
$P^{1/2}$ & $C_2^{3S1}$\\
$P^{3/2}$ & $C_2^{3P0}$, $C_2^{3P2}$\\
$P^2$ & Renormalized ``leading" two-pion exchange $V_{2 \pi}^{(2)}$, $C^{1P1}$, $C^{3P1}$. $C_4^{1S0}$, $C^{\epsilon1}$\\
$P^{5/2}$ & $C_4^{3S1}$\\
$P^3$ & Renormalized ``sub-leading" two-pion exchange, $V_{2 \pi}^{(3)}$\\
\hline
\end{tabular}
\caption{Order of different pieces of the potential $V$ in $\chi$EFT, if $\Lambda_{NN}$ is identified as a low-energy scale. Adapted from Ref.~\cite{Bi09}.}
\end{center}
\label{table-birse}
\end{table}

\begin{figure}[thbp]
\centerline{
\includegraphics[width=5.0in]{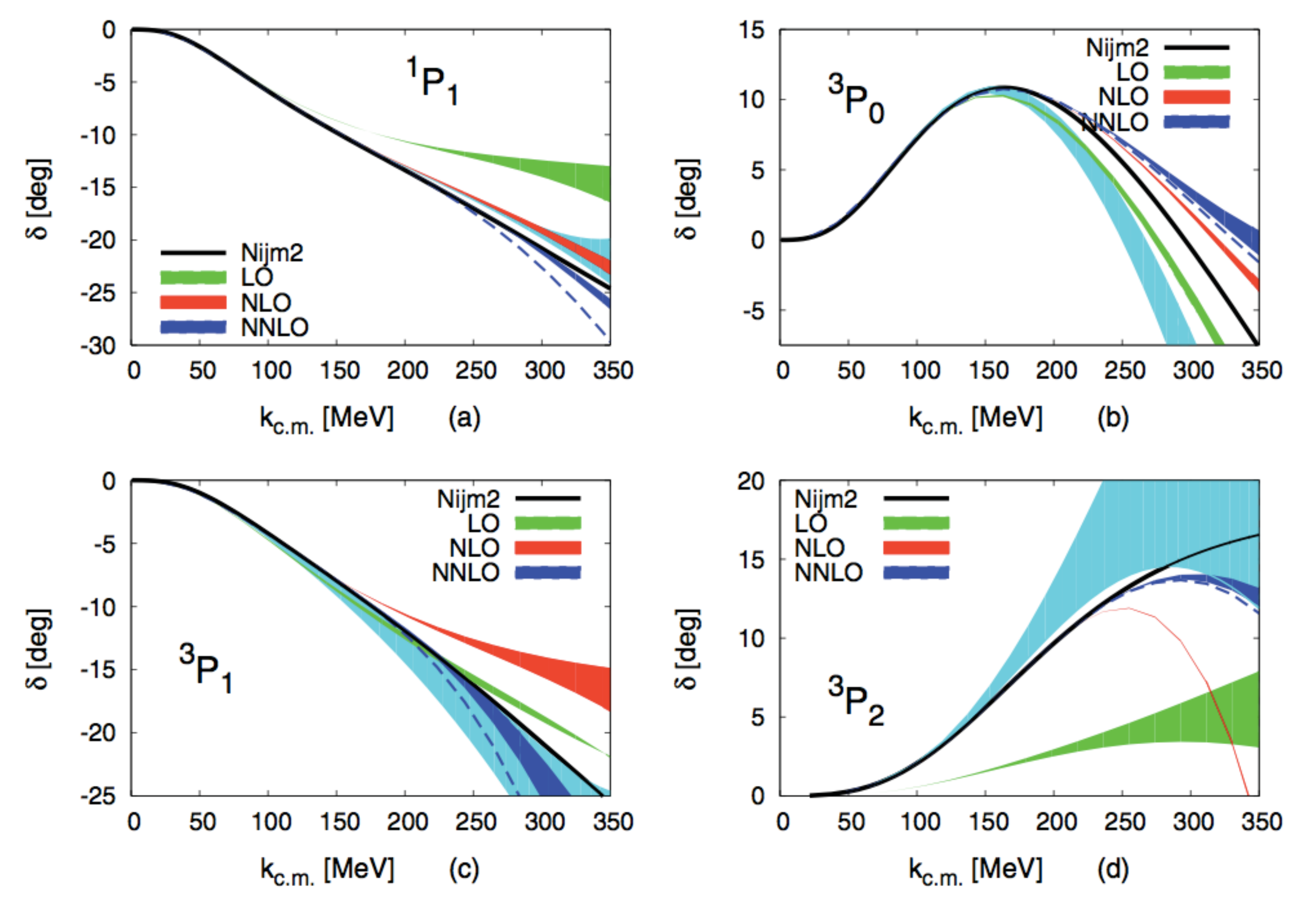}}
\caption{Phase shifts in different NN P-wave channels as a function of lab energy. The bands indicate the variation in phase shifts with variation in the cutoff $r_c$ at different orders in the perturbative expansion discussed here. The black line can be taken as a parameterization of experimental data.
The light-blue shaded region shows the result obtained with the NNLO $\chi$PT potential~\cite{Ep99} in the ``Weinberg" power counting usually used in $\chi$EFT calculations. Taken from Ref.~\cite{PV11}. Used with permission.}
\label{fig-PwavephasesNNLO}
\end{figure}

Turning first to the controversial P-waves, Fig.~\ref{fig-PwavephasesNNLO} displays results from Ref.~\cite{PV11}. The bands are generated by varying the cutoff $r_c$ between 0.6 and 0.9 fm, with the light-blue band obtained by solving the Schr\"odinger equation with the NNLO $\chi$PT potential. The large width of that band shows that the lack of renormalization displayed at LO in Ref.~\cite{NTvK} persists at NNLO, especially in the ${}^3$P$_2$ phase shift. In contrast, the ``new leading order" discussed in Sec.~\ref{sec-newLO} is given by the green band, and is stable with respect to $r_c$. 
The red and dark blue bands are generated via a perturbative treatment of, respectively, the leading (from $V^{(2)}$) and sub-leading (from $V^{(3)}$), two-pion-exchange corrections. Results are not completely stable with respect to $r_c$, but the analysis of Sec.~\ref{sec-subleading} suggests they will not be, until another contact interaction is added to the calculation.

I will not display results for the ${}^3$S$_1$-${}^3$D$_1$ channel here. Suffice to say, even though here the new LO is the same as the LO defined by Eq.~(\ref{eq:V0}) and (\ref{eq:SE}), higher-order corrections in the ``new" and ``standard" $\chi$EFT expansion behave rather differently. As for the ${}^3$P$_2$-${}^3$F$_2$ channel, a perturbative treatment of $V_{2 \pi}^{(2)}$ and $V_{2 \pi}^{(3)}$ yields results that are quite stable over a significant range of $r_c$. A non-perturbative solution of the Schr\"odinger equation with $V$ summed up to $V^{(3)}$ does not.

\begin{figure}[hbtp]
\centerline{
\includegraphics[width=3in]{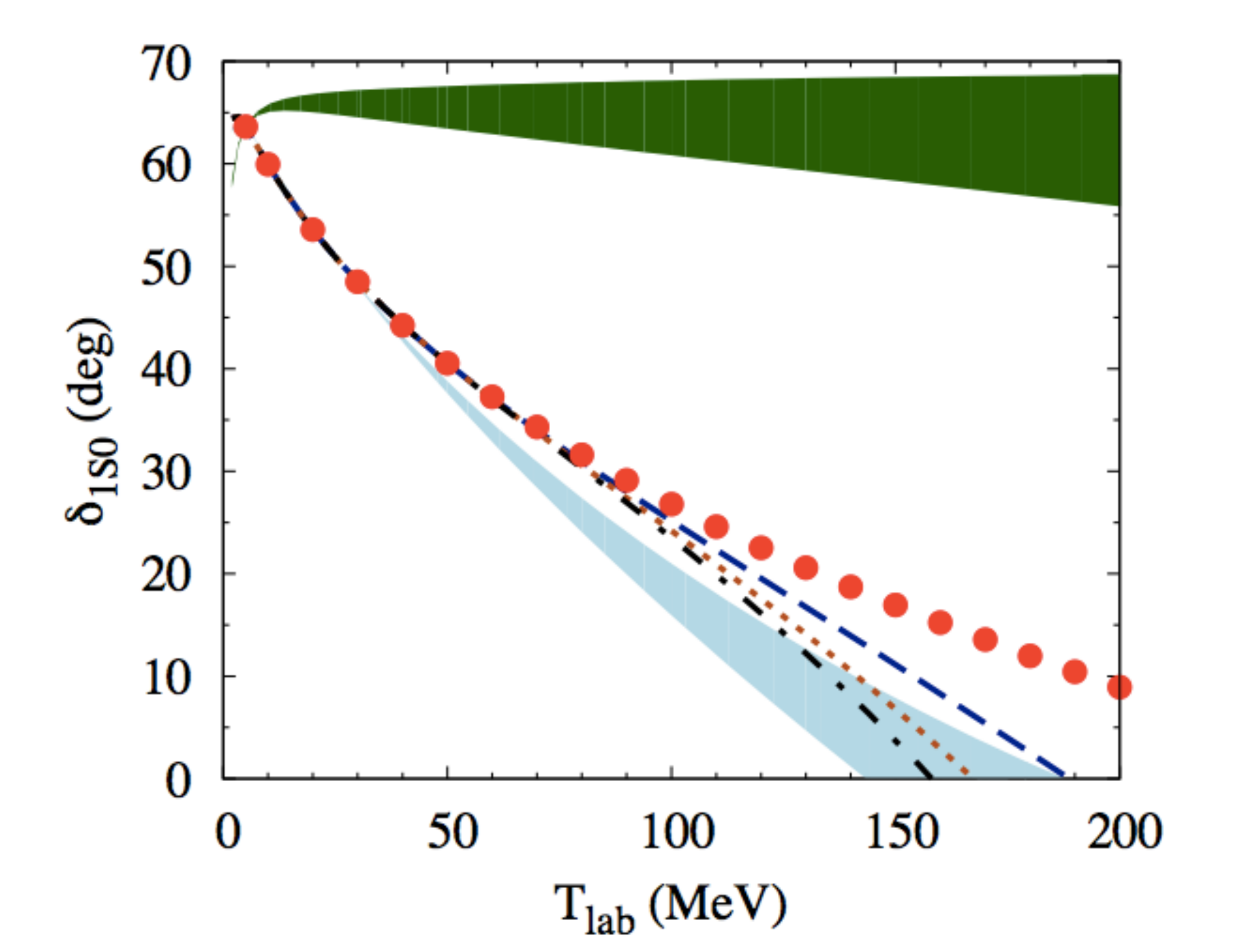}}
\caption{Phase shifts in the ${}^1$S$_0$ channel with perturbative treatment of higher-order effects at $O(P^{-1})$ (green band), $O(P^1)$ (light blue band), and $O(P^2)$ (dashed, dotted, and dot-dashed lines). In each case the momentum-space cutoff is varied between 0.5--2 GeV. The red dots are from the Nijmegen PWA. Taken from Ref.~\cite{LY12}. Used with permission.} 
\label{fig-1S0}
\end{figure}

The other prominent NN channel where the ``standard" $\chi$EFT prescription of inserting $V$ into the Schr\"odinger equation does not lead to stable results in this range of $r_c$ is the ${}^1$S$_0$. 
As discussed in Sec.~\ref{sec-subleading}, and demonstrated numerically in Refs.~\cite{Bi10,PV09,PV11,LY12}, additional ${}^1$S$_0$ contact terms are needed in order to yield stable results beyond LO. Fig.~\ref{fig-1S0} shows a calculation from Ref.~\cite{LY12}, organized according to the power counting shown in Table~\ref{table-birse}. Good agreement is found at $O(P^2)$.

\section{Omissions and Outlook}

\label{sec-conc}

Recent progress in $\chi$EFT for the NN system has been significant. The systematic inclusion of perturbative corrections to the ``new leading order" of Ref.~\cite{NTvK} is leading to results that maintain renormalization-group invariance over a wide range of cutoffs, and achieve good agreement with data at $O(P^3)$ relative to leading. 

Nevertheless, unresolved issues remain. There is disagreement about the power counting of short-distance operators in partial waves where one-pion exchange is repulsive, e.g. the ${}^3$P$_1$. It is also, as yet, unclear how many contact terms are needed to stabilize the ${}^3$P$_2$-${}^3$F$_2$ partial wave at a given order in the expansion. Meanwhile, some authors (e.g. Refs.~\cite{NTvK,PV11}) treat one-pion exchange non-perturbatively in D-waves, leading to even more marked differences with the standard $\chi$PT counting than in P-waves.
Whether such a treatment is necessary is unclear, as the D-wave phase shifts are quite small, and a perturbative treatment of the $\chi$PT NN potential yields moderately good agreement with data~\cite{Ka97}. 
Another point warranting further research is the extent to which the fine tuning in the ${}^3$S$_1$ channel affects contact operators there. Birse argues that it enhances 
${}^3$S$_1$ operators further over the order given by the analysis in Sec.~\ref{sec-subleading}~\cite{Bi06,Bi09}. But, e.g., Pavon Valderramma does not include this additional enhancement in his analysis, and still manages to describe the ${}^3$S$_1$-${}^3$D$_1$ phase shifts well over a wide range of cutoffs~\cite{PV09}. Finally, there is the question of what is leading order: $O(P^{-1})$, $O(P^{-1/2})$, $O(P^0)$? This last may seem an academic question, but it is actually critically connected to our understanding of what physics drives the non-perturbative nature of the NN system. Is it the tensor one-pion exchange, a fine-tuned contact interaction, or something else? 

Since I am confessing my sins of omission I add that this contribution did not have space to do justice to {\it all} recent work on the NN system in $\chi$EFT. Particularly notable is Aldaladejo and Oller application of N/D methods in order to remove the right-hand cut, which allows them to power count an object with only left-hand cuts~\cite{AO12}. Gasparyan {\it et al.} are also marrying $\chi$PT and dispersion relations in order to attack the NN problem~\cite{Ga12}.
And Szpigel and Timoteo have applied subtractive renormalization in the ${}^1$S$_0$ and proposed a power counting different to both the standard one and that of Table~\ref{table-birse}~\cite{ST12}. More generally, the bibliography I provide is, of necessity, partial---in both senses of the word. I apologize in advance to those whose work is not represented, but this is a conference proceeding, not a review paper, and  my goal was to provide sufficient references to permit interested parties to trace ideas and details back to original papers.

I see the work described here as releasing the few-nucleon portion of the ``Chiral Dynamics" community  from the state of double-mindedness~\cite{Jas} in which it has existed for the last two meetings in this series. At both these meetings we knew that the $\chi$PT expansion for the nuclear force worked well, but also knew that the way it was being applied did not constitute a properly renormalized EFT. In the last three years significant work by a talented group of young researchers has shown that there {\it is} a $\chi$EFT expansion for the NN force which is (a) perturbative for all corrections beyond leading order; (b) demonstrably renormalization-group invariant over a wide range of cutoffs at each order; and (c) provides a good description of NN phase shifts. It is my hope and expectation that future work on $\chi$EFT in the NN system will resolve the remaining technical issues surrounding this new $\chi$EFT expansion, and also assess the extent to which the inclusion of the $\Delta(1232)$ as an explicit degree of freedom improves the description of NN data and the convergence of the theory. The resulting, truly systematic, picture of chiral dynamics will have implications for three-nucleon forces and electroweak operators, and these will have to be teased out. Only then will the successes listed in the Introduction and the beautiful results presented at this conference on, e.g. isospin violation in $\pi$d~\cite{Ho12}, threshold M1 capture~\cite{Gi10}, weak capture in deuterium~\cite{Ma12}, and magnetic moments of light nuclei~\cite{Pa12} stand on a completely firm EFT foundation. The tools to build that foundation are now well understood, and 
its construction need not mar the notable phenomenological success of $\chi$EFT for few-nucleon systems. 

\acknowledgments

This work was supported by the U.~S.~Department of Energy under grant no. DE-FG02-93ER40756. I thank the organizers of CD12 for inviting me to an excellent meeting. I
 am also grateful to my interlocutors, M.~C.~Birse, E.~Epelbaum, J.~Gegelia, B.~Long, A.~Nogga, M.~Pavon Valderrama,  U.~van Kolck, and C.-J.~Yang for sharing their insights into this subject. I emphasize, however, that any misrepresentation of their work contained herein is entirely my responsibility.


\begin{thebibliography}{99}
\bibitem{We90} S.~Weinberg, 
Phys.\ Lett.\ B \textbf{251}, 288 (1990); Nucl.\ Phys.\ B \textbf{363}, 3 (1991).

\bibitem{Or96} C.~Ordonez, L.~Ray and U.~van Kolck, 
Phys.\ Rev.\ C \textbf{53}, 2086 (1996).

\bibitem{Ep99} E.~Epelbaum, W.~Gl\"ockle, and U.-G.~Mei\ss ner, 
Nucl.\ Phys.\ A \textbf{671}, 295 (2000).

\bibitem{EM02}
  D. R. Entem and R. Machleidt
  Phys.\ Lett.\ B {\bf 524}, 93 (2002).


\bibitem{EM03} D.~R.~Entem and R.~Machleidt, 
Phys.\ Rev.\ C \textbf{68}, 041001(R) (2003).

\bibitem{Ep05} E.~Epelbaum, W.~Gl\"ockle and U-G.~Mei\ss ner, 
Nucl.\ Phys.\ A \textbf{747}, 362 (2005).


  \bibitem{vK94}
  U.~van Kolck,
  Phys.\ Rev.\  C {\bf 49}, 2932 (1994).

  
  \bibitem{Ep02}
  E.~Epelbaum {\it et al.},
  Phys.\ Rev.\  C {\bf 66}, 064001 (2002).


\bibitem{Be02} S.~R.~Beane, P.~F.~Bedaque, M.~J.~Savage, and U.~van Kolck, 
Nucl.\ Phys.\ A \textbf{700}, 377 (2002).

\bibitem{ES03}
  D.~Eiras and J.~Soto,
  Eur.\ Phys.\ J.\  A {\bf 17}, 89 (2003).

\bibitem{NTvK} A.~Nogga, R.G.E.~Timmermans and U.~van Kolck, 
Phys.\ Rev.\ C \textbf{72}, 054006 (2005).

\bibitem{Bi06}
  M.~C.~Birse,
  Phys.\ Rev.\  C {\bf 74}, 014003 (2006).
 
\bibitem{PVRA06}
M.~Pavon Valderrama and E. Ruiz Arriola,
  Phys.\ Rev.\  C {\bf 74}, 064004 (2006).

\bibitem{Ya08}
  C.-J.~Yang, Ch.~Elster and D.~R.~Phillips
  Phys.\ Rev.\ C {\bf 77}, 014002 (2008).  


\bibitem{Le97} G.~P.~Lepage, 
arXiv:nucl-th/9706029. 


\bibitem{EM06} E.~Epelbaum and Ulf-G.~Mei\ss ner, 
arXiv:nucl-th/0609037. 


\bibitem{MVN12}
  P.~Maris, J.~P.~Vary and P.~Navratil,
  Phys.\ Rev.\ C {\bf 87}, 014327 (2013).

\bibitem{Ro12} See R.~Roth, these proceedings, and references therein.

\bibitem{KN11}
  N.~Kalantar-Nayestanaki, E.~Epelbaum, J.~G.~Messchendorp and A.~Nogga,
  Rept.\ Prog.\ Phys.\  {\bf 75}, 016301 (2012).

\bibitem{NQ11}
  P.~Navratil
   and S.~Quaglioni,
  Phys.\ Rev.\ Lett.\  {\bf 108}, 042503 (2012).

\bibitem{Ko12} S.~K\"olling, these proceedings, and references therein. 

\bibitem{Pa12} S.~Pastore, these proceedings, and references therein.

\bibitem{QN07}
S.~Quaglioni and P.~Navratil, Phys. Lett. {\bf B652}, 370 (2007).

\bibitem{Ma12} L.~E.~Marcucci, A.~Kievsky, S.~Rosati, R.~Schiavilla and M.~Viviani,
  Phys.\ Rev.\ Lett.\  {\bf 108}, 052502 (2012),
  and  L.~E.~Marcucci,, these proceedings, and references therein.

\bibitem{Gr12}  H.~W.~Grie\ss hammer, J.~A.~McGovern, D.~R.~Phillips and G.~Feldman,
  Prog.\ Part.\ Nucl.\ Phys.\  {\bf 67}, 841 (2012).


\bibitem{Ya09A} C.-J. Yang, Ch. Elster and D. R. Phillips, Phys.\ Rev.\ C 
\textbf{80}, 034002 (2009).

\bibitem{Ya09B} C.-J. Yang, Ch. Elster and D. R. Phillips, 
  Phys.\ Rev.\ C {\bf 80}, 044002 (2009).
  
 
     \bibitem{Ka98}
  D.~B.~Kaplan, M.~J.~Savage and M.~B.~Wise,
  Phys.\ Lett.\  B {\bf 424}, 390 (1998).
  Nucl.\ Phys.\  B {\bf 534}, 329 (1998).

    

  
\bibitem{FMS}
  S.~Fleming, T.~Mehen and I.~W.~Stewart,
  Nucl.\ Phys.\ A {\bf 677}, 313 (2000).

 
\bibitem{Ka97} N.~Kaiser, R.~Brockmann and W.~Weise, 
Nucl.\ Phys.\ A \textbf{625}, 758 (1997). 

   
\bibitem{PVRA05}
  M.~Pavon Valderrama and E.~Ruiz Arriola,
  Phys.\ Rev.\  C {\bf 72}, 054002 (2005).
    

\bibitem{PV08}
  M.~Pavon Valderrama, A.~Nogga, E.~Ruiz Arriola and D.~R.~Phillips,
  Eur.\ Phys.\ J.\  A {\bf 36}, 315 (2008).



\bibitem{Case50}
  K.~M.~Case,
  Phys.\ Rev.\  {\bf 80}, 797 (1950).

\bibitem{Sprung94}
  D.~W.~L.~Sprung~{\it et al.},
  Phys.\ Rev.\ C {\bf 49}, 2942 (1994).

\bibitem{Be01}
  S.~R.~Beane~{\it et al.},
  Phys.\ Rev.\ A {\bf 64}, 042103 (2001).
  
  \bibitem{BKV}
  S.~R.~Beane, D.~B.~Kaplan and A.~Vuorinen,
  Phys.\ Rev.\ C {\bf 80}, 011001 (2009).


\bibitem{GE12} 
  E.~Epelbaum and J.~Gegelia,
  Phys.\ Lett.\ B {\bf 716}, 338 (2012), and 
  arXiv:1210.3964 [nucl-th].
  
 

\bibitem{Sh08}
  D.~Shukla, D.~R.~Phillips and E.~Mortenson,
  J.\ Phys.\ G {\bf 35}, 115009 (2008).

\bibitem{Bi10}
  M.~C.~Birse,
  Eur.\ Phys.\ J.\ A {\bf 46}, 231 (2010).

\bibitem{PV09}
    M.~Pavon Valderrama,
  Phys.\ Rev.\ C {\bf 83}, 024003 (2011).

\bibitem{PV11}
  M.~Pavon Valderrama,
  Phys.\ Rev.\ C {\bf 84}, 064002 (2011).

\bibitem{LY12}
  B.~Long and C.~J.~Yang,
  Phys.\ Rev.\ C {\bf 86}, 024001 (2012).
  
  \bibitem{vK99}  U.~van Kolck,
  Nucl.\ Phys.\  A {\bf 645}  (1999) 273.   
  
   \bibitem{Ge98}
  J.~Gegelia,
  Phys.\ Lett.\  B {\bf 429} , 227 (1998).
  
\bibitem{Bi99}
  M.~C.~Birse, J.~A.~McGovern and K.~G.~Richardson,
  Phys.\ Lett.\  B {\bf 464}, 169 (1999).

    
\bibitem{Bi09} M.~C.~Birse, PoS (CD09), 078.
  
\bibitem{LY11A}
  B.~Long and C.~J.~Yang,
  Phys.\ Rev.\ C {\bf 84}, 057001 (2011).

\bibitem{LY11B}
  B.~Long and C.~J.~Yang,
  Phys.\ Rev.\ C {\bf 85}, 034002 (2012).
  
  


\bibitem{Long}
B.~Long, these proceedings.

\bibitem{Pavon}
M.~Pavon Valderrama, these proceedings.

\bibitem{AO12}
  M.~Albaladejo and J.~A.~Oller,
  Phys.\ Rev.\ C {\bf 86}, 034005 (2012).

 
 \bibitem{Ga12}
    A.~M.~Gasparyan, M.~F.~M.~Lutz and E.~Epelbaum,
  arXiv:1212.3057 [nucl-th], and these proceedings.
  
  
\bibitem{ST12}
  S.~Szpigel and V.~S.~Timoteo,
  J.\ Phys.\ G {\bf 39}, 105102 (2012).

\bibitem{Jas}
James 1:8, New International Version.

\bibitem{Ho12}   M.~Hoferichter~{\it et al.}, these proceedings,
  arXiv:1211.1145 [nucl-th].

\bibitem{Gi10}   L.~Girlanda {\it et al.},
  Phys.\ Rev.\ Lett.\  {\bf 105}, 232502 (2010),
and L.~Girlanda, these proceedings.
\end{thebibliography}
\end{document}